# Аналитическое исследование масштабируемости фреймворка «мастер-рабочие» применительно к многопроцессорным системам с распределенной памятью[*]

Л.Б. Соколинский

Южно-Уральский государственный университет

Работа посвящена аналитическому исследованию масштабируемости фреймворка организации параллельных вычислений по схеме «мастер-рабочие» применительно к многопроцессорным системам с распределенной памятью. В статье строится новая модель параллельных вычислений *BSF*, основанная на моделях *BSP*, и *SPMD*. Модель *BSF* ориентирована на вычислительные задачи с высокой временной сложностью. Определяется архитектура *BSF*-компьютера и описывается структура *BSF*-программы. Описывается формальная стоимостная метрика, с помощью которой получаются верхние оценки масштабируемости параллельных программ, основанных на модели *BSF*, применительно к вычислительным системам с распределенной памятью. Также выводятся формулы для оценки эффективности распараллеливания *BSF*-программ.

*Ключевые слова:* параллельное программирование, фреймворк «мастер-рабочие», модель *BSF*, временная сложность, масштабируемость, многопроцессорные системы с распределенной памятью.

## 1. Введение

Суперкомпьютер TaihuLight с массово-параллельной архитектурой, занимающий первое место в списке TOP-500 [1] самых мощных суперкомпьютеров мира (ноябрь 2016), имеет 40 960 процессорных узлов, каждый из которых включает в себя 260 процессорных ядер. Общая оперативная память системы составляет 1.3 Петабайт, пиковая производительность превышает 120 петафлопс. Анализ динамики роста производительности суперкомпьютеров (см. рис. 1) показывает, что через 8-9 лет самый мощный суперкомпьютер становится рядовой системой, и что через 5-6 лет мы можем ожидать появление суперкомпьютера с экзафлопным уровнем производительности. Появление столь мощных многопроцессорных вычислительных систем выдвигает на первый план вопросы, связанные с разработкой фреймворков (шаблонов), позволяющих создавать высокомасштабируемые параллельные программы, ориентированные на системы с распределенной памятью. При этом особенно важной является проблема разработки моделей параллельных вычислений, позволяющих на ранней стадии проектирования программы оценить ее масштабируемость.

В данной работе предлагается новая модель параллельных вычислений *BSF* (*Bulk Synchronous Farm*) – блочно-синхронная ферма, основанная на моделях «мастер-рабочие», *BSP*, и *SPMD*. Модель *BSF* ориентирована на вычислительные системы с массовым параллелизмом на распределенной памяти, включающие в себя сотни тысяч процессорных узлов, и имеющие экзафлопный уровень производительности. Модель *BSF* включает в себя каркас (skeleton) для разработки параллельных программ и стоимостную метрику для оценки масштабируемости приложения.

Статья имеет следующую структуру. В разделе 2 дается краткий обзор концептуальных моделей параллельных вычислений, лежащих в основе новой модели *BSF*. В разделе 3 приводятся общие требования к модели вычислений. Раздел 4 посвящен описанию модели *BSF*. В разделе 5 строится стоимостная метрика, и даются аналитические оценки масштабируемости *BSF*-приложений, а также исследуется вопрос эффективности распараллеливания *BSF*-программ. В разделе 6 суммируются полученные результаты и намечаются направления дальнейших исследований.

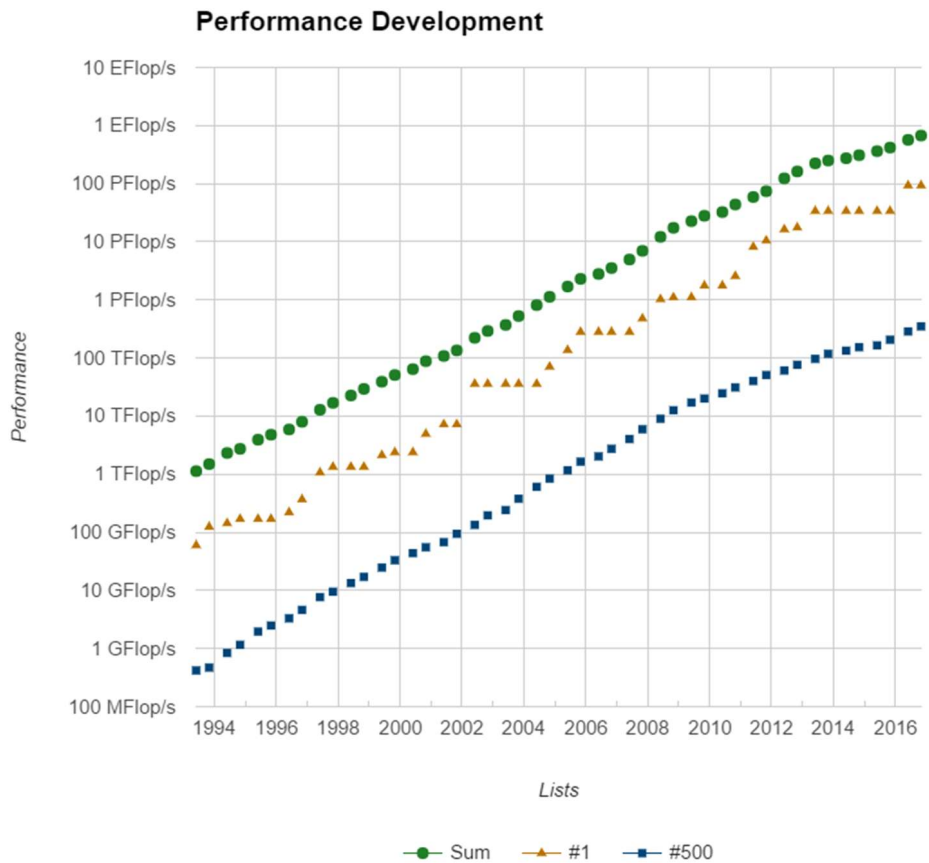

**Рис. 1**. Динамика роста производительности суперкомпьютеров в TOP500.

## 2. Модели параллельных вычислений

Одним из популярных фреймворков, используемых в параллельном и распределенном программировании, является *парадигма «мастер-рабочие»* [2–4]. В соответствии с этой парадигмой процессорные узлы вычислительной системы делятся на два множества: узлы-мастера и узлы-рабочие. Задача, решаемая на многопроцессорной системе, разбивается на независимые подзадачи. Каждая подзадача, в свою очередь, разбивается на три последовательные стадии: стадия предобработки, стадия вычислений и стадия постобработки. Стадии предобработки и постобработки выполняются узлами-мастерами, стадии вычислений выполняются узлами-рабочими. В каждый момент времени любой процессорный узел может выполнять только одну стадию. Предполагается, что количество подзадач совпадает с количеством узлов-рабочих и превышает количество узлов-мастеров. Решение задачи происходит следующим образом. Очередная подзадача назначается некоторому узлу-мастеру, который выполняет стадию предобработки. После этого он посылает задание (передает данные) узлу-рабочему, который должен выполнить стадию вычислений указанной подзадачи. После того, как узел-рабочий завершил требуемые вычисления, он посылает отчет (передает данные) тому узлу-мастеру, от которого получил задание. На этом выполнение подзадачи заканчивается. Задача считается выполненной целиком, когда выполнены все ее подзадачи. Фреймворк «мастер-рабочий» очень часто используется в параллельном программировании при реализации различных приложений, ориентированных на многопроцессорные системы с распределенной памятью (см., например, [5–10]). При этом наиболее популярна конфигурация, включающая одного мастера и множество рабочих (см. рис. 2). Основной проблемой модели «мастер-рабочие» является нахождение такого расписания вычислений, при котором время выполнения задачи будет минимальным. Известно, что указанная проблема является в общем случае NP-сложной [11]. В определенной мере эту проблему можно решить, используя модель *SPMD*.

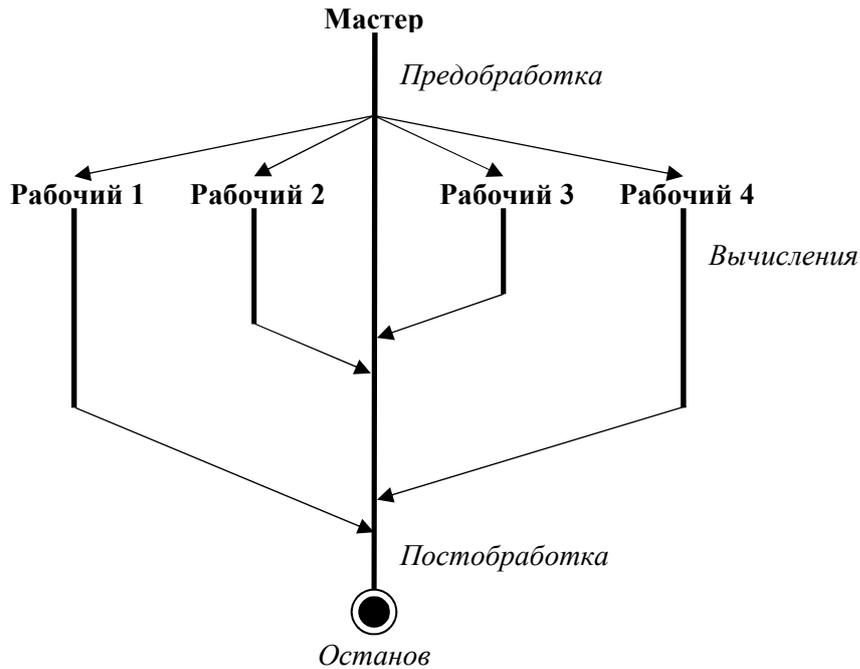

**Рис. 2**. Фреймворк «мастер-рабочие».
Полужирными линиями обозначены потоки вычислений,
тонкими линями со стрелками – потоки данных.

*SPMD* (*Single Program Multiple Data*) [3, 12] – популярная парадигма параллельного программирования, в соответствии с которой все процессорные узлы выполняют одну и ту же программу, но обрабатывают различные данные. Какие именно данные необходимо обрабатывать тому или иному процессорному узлу, определяется его уникальным номером, который является параметром программы. Данный подход наиболее часто используется в сочетании с технологией MPI (Message Passing Interface) [13], которая де-факто является стандартом для параллельного программирования на распределенной памяти.

*Модель BSP* (*Bulk-Synchronous Parallelism*) была предложена Валиантом (Valiant) в работе [14]. Данная модель широко используется при разработке и анализе параллельных алгоритмов и программ. *BSP-компьютер* представляет собой систему из $p$ процессоров, имеющих приватную память, и соединенных сетью, позволяющей передавать пакеты данных фиксированного размера от одного процессора другому. Для соединительной сети вводятся следующие характеристики: $g$ – время, необходимое для передачи по сети одного пакета; $L$ – время, необходимое для инициализации передачи данных от одного процессора другому.

*BSP-программа* состоит из $p$ потоков команд, каждый из которых назначается отдельному процессору, и делится на *супершаги*, которые выполняются последовательно относительно друг друга. Каждый *супершаг*, в свою очередь, включает в себя следующие четыре последовательных шага: 1) вычисления на каждом процессоре с использованием только локальных данных; 2) глобальная барьерная синхронизация; 3) пересылка данных от любого процессора любым другим процессорам; 4) глобальная барьерная синхронизация. Переданные данные становятся доступными для использования только после барьерной синхронизации. Пример *BSP-программы* приведен на рис. 3.

*Стоимостная функция* в модели *BSP* строится следующим образом [15]. Пусть *BSP-программа* состоит из $S$ супершагов. Пусть $w_i$ – максимальное время, затраченное каждым процессором на локальные вычисления, $h_i$ – максимальное количество пакетов, посланных или

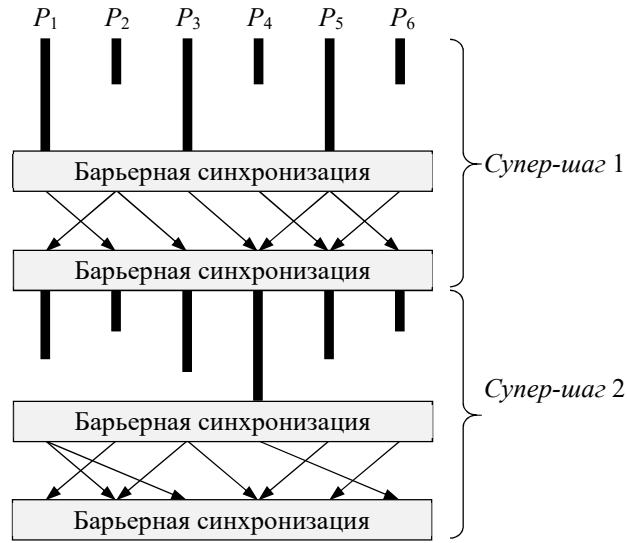

**Рис. 3**. *BSP*-программа из двух супершагов на шести процессорах.
Жирными линиями обозначены локальные вычисления,
тонкими линями со стрелками – пересылка данных.

полученных каждым процессором, на *i*-том супершаге. Тогда общее время $t_i$, затрачиваемое системой на выполнение *i*-того супершага, вычисляется по формуле

$$t_i = w_i + gh_i + L .$$

Время $T$ выполнения всей программы определяется по формуле

$$T = W + Hg + LS , \qquad (1)$$

где $W = \sum_{i=1}^{S} w_i$ и $H = g\sum_{i=1}^{S} h_i$ .

Для того, чтобы облегчить программистам использование той или иной модели параллельных вычислении на практике, применяются *программные каркасы* [16]. Одним из наиболее популярных является *ферменный каркас* (*farm skeleton*), реализующий фреймворк «мастер-рабочие» [17]. Такой каркас представляет собой программную структуру, полностью реализующую фреймворк «мастер-рабочие», однако вместо эффективного кода и реальных данных она содержит заглушки, которые должны быть заменены по определенным правилам на фрагменты кода, реализующие целевую задачу.

## 3. Требования к модели параллельных вычислении

Модель параллельных вычислений в общем случае должна включать в себя следующие четыре компонента, некоторые из которых в определенных случаях могут быть тривиальны [18].
1. *Архитектурный компонент*, описываемый как помеченный граф, узлы которого соответствуют модулям с различной функциональностью, а дуги – межмодульным соединениям для передачи данных.
2. *Спецификационный компонент*, определяющий, что есть корректная программа.
3. *Компонент выполнения*, определяющий, как взаимодействуют между собой архитектурные модули при выполнении корректной программы.
4. *Стоимостный компонент*, определяющий одну или более стоимостных метрик для оценки времени выполнения корректной программы.

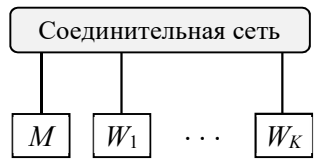

**Рис. 4**. *BSF*-компьютер. M – мастер;
$W_1, …, W_K$ – рабочие.

В качестве наиболее важных свойств модели параллельных вычислений обычно выделяют следующие [19].

– *Юзабилити*, определяющая легкость описания алгоритма и анализа его стоимости средствами модели (модель должна быть легкой в использовании).
– *Адекватность*, выражающаяся в соответствии реального времени выполнения программ и временной стоимости, полученной аналитически с помощью стоимостных метрик модели (программа, имеющая меньшую временную стоимость должна выполняться быстрее).
– *Портируемость*, характеризующая широту класса целевых платформ, для которых модель оказывается применимой.

## 4. Модель *BSF*

*Модель параллельных вычислений BSF* (*Bulk Synchronous Farm*) – *блочно-синхронная ферма* ориентирована на многопроцессорные системы с кластерной архитектурой и архитектурой MPP. *BSF-компьютер* представляет собой множество однородных процессорных узлов с приватной памятью, соединенных сетью, позволяющей передавать данные от одного процессорного узла другому. Среди процессорных узлов выделяется один, называемый *узлом-мастером* (или кратко *мастером*). Остальные $K$ узлов называются *узлами-рабочими* (или просто *рабочими*). В *BSF*-компьютере должен быть по крайней мере один узел мастер и один рабочий ($K \geq 1$). Схематично архитектура *BSF*-компьютера изображена на рис. 4.

*BSF*-компьютер работает по схеме *SPMD*. *BSF-программа* состоит из последовательности *супершагов* и глобальных барьерных синхронизаций, выполняемых мастером и всеми рабочими. Каждый супершаг делится на секции двух типов: *секции мастера*, выполняемые только мастером, и *секции рабочего*, выполняемые только рабочими. Относительный порядок секций мастера и рабочего в рамках супершага не существен. Данные, обрабатываемые конкретным узлом-рабочим, определяются его номером, являющимся параметром среды исполнения.

*BSF*-программа включает в себя следующие последовательные *разделы* (см. рис. 5):

- инициализация;
- итерационный процесс;
- завершение.

*Инициализация* представляет собой супершаг, в ходе которого мастер и рабочие считывают или генерируют исходные данные. Инициализация завершается барьерной синхронизацией. *Итерационный процесс* состоит в многократном повторении *тела итерационного процесса* до тех пор, пока не выполнено условие выхода, проверяемое мастером. В разделе *завершение* осуществляется вывод или сохранение результатов и завершение программы.

*Тело итерационного процесса* включает в себя следующие супершаги:

1) передача рабочим заданий от мастера;
2) выполнение задания (рабочими);
3) передача мастеру результатов от рабочих;
4) обработка полученных результатов мастером.

На первом супершаге мастер рассылает всем рабочим одинаковые задания. На втором супершаге происходит выполнение полученного задания рабочими (мастер при этом простаивает). Все рабочие выполняют один и тот же программный код, но обрабатывают различные данные, адреса

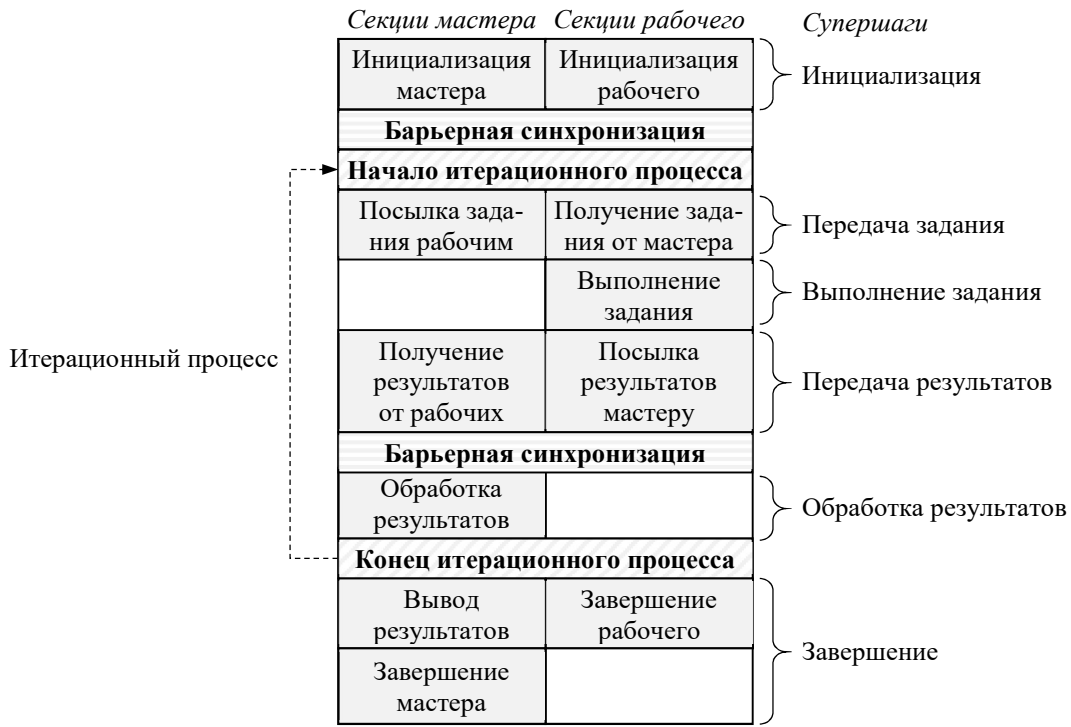

**Рис. 5**. Структура *BSF*-программы.

которых определяются по номеру рабочего. Это означает, что все рабочие тратят на вычисления одно и то же время. Никаких пересылок данных при выполнении задания не происходит. Это является важным свойством модели *BSF*. На третьем супершаге все рабочие пересылают мастеру полученные результаты. Суммарный объем результатов является атрибутом задачи и не зависит от количества рабочих. После этого происходит глобальная барьерная синхронизация. В ходе четвертого супершага мастер производит обработку и анализ полученных результатов. Рабочие в это время простаивают. Время обработки мастером результатов, полученных от рабочих, является параметром задачи и не зависит от количества рабочих. Если после обработки результатов условие выхода оказывается истинным, то происходит выход из итерационного процесса, в противном случае осуществляется переход на первый супершаг итерационного процесса. На четвертом супершаге происходит вывод или сохранение результатов и завершение работы мастера и рабочих. Графическая иллюстрация работы *BSF*-программы приведена на. рис. 6.

*Областью применения* модели *BSF* являются *масштабируемые* итерационные численные методы, имеющие высокую вычислительную сложность итерации при относительно невысокой стоимости коммуникаций. Под *масштабируемым итерационным методом* понимается метод, допускающий разбиение итерации на подзадачи, не требующие обменов данными. Пример такого метода можно найти в работе [20].

## 5. Исследование масштабируемости модели *BSF*

Основной характеристикой масштабируемости является ускорение, вычисляемое как отношение времени выполнения программы на одном процессорном узле ко времени выполнения на $K$ узлах. В данном разделе мы проведем аналитическое исследование масштабируемости модели *BSF*. Для этого нам понадобится оценка временных затрат на выполнение *BSF*-программы.

Мы предполагаем, что временные затраты на инициализацию и завершение *BSF*-программы пренебрежимо малы по сравнению с затратами на выполнение итерационного процесса. Стоимость итерационного процесса получается как сумма стоимостей отдельных итераций. Поэтому для оценки времени выполнения *BSF*-программы нам достаточно получить оценку временной стоимости одной итерации.

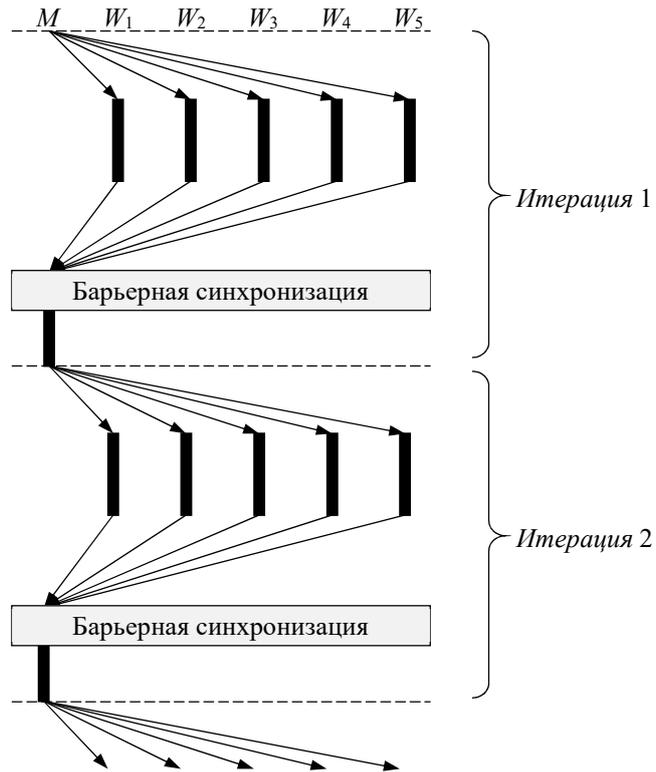

**Рис. 6.** Иллюстрация работы *BSF*-программы
с одним мастером $M$ и пятью рабочими $W_1,\ldots,W_5$
(жирными линиями обозначены локальные вычисления,
тонкими линиями со стрелками – пересылка данных,
пунктирными линиями – границы одной итерации).

Рассмотрим сначала конфигурацию вычислительной системы в составе мастера и одного рабочего. Пусть $t_s$ – время, необходимое для посылки задания рабочему (без учета латентности); $t_r$ – время, необходимое для передачи результата мастеру от рабочего (без учета латентности); $t_p$ – время обработки мастером результатов, полученных от рабочего; $L$ – затраты на инициализацию операции передачи сообщения (латентность); $t_w$ – время выполнения всех вычислений в рамках итерации одним рабочим (в ситуации, когда задача решается бригадой из одного рабочего). Общее время $T_1$ выполнения итерации одним мастером и бригадой из одного рабочего может быть вычислено следующим образом:

$$T_1 = t_s + t_w + L + t_p + t_r + L, \qquad (2)$$

что равносильно

$$T_1 = 2L + t_s + t_r + t_p + t_w. \qquad (3)$$

Теперь рассмотрим конфигурацию вычислительной системы в составе одного мастера и $K$ рабочих. Все рабочие получают от мастера одно и то же сообщение, поэтому общее время передачи сообщений от мастера рабочим составит $K(L+t_s)$. Все рабочие выполняют один и тот же код над своей частью данных, поэтому время выполнения всех вычислений $K$ рабочими в рамках одной итерации будет равно $t_w/K$. Суммарный объем результатов, вычисленных рабочими, является параметром задачи и не зависит от $K$, поэтому общее время передачи сообщений мастеру от рабочих составит $K \cdot L + t_r$. Время обработки мастером результатов, полученных от рабочих, также является параметром задачи и не зависит от количества рабочих. Таким образом, общее время $T_K$ выполнения итерации в системе с одним мастером и $K$ рабочими может быть вычислено следующим образом:

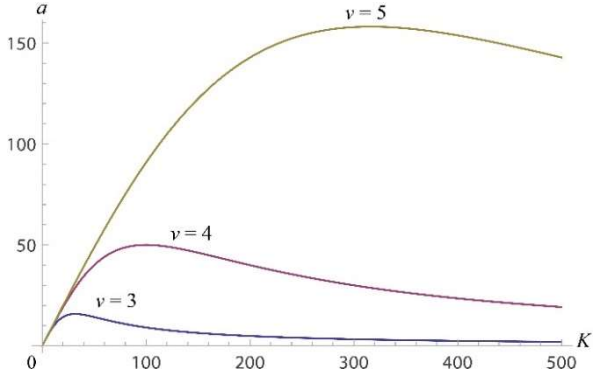 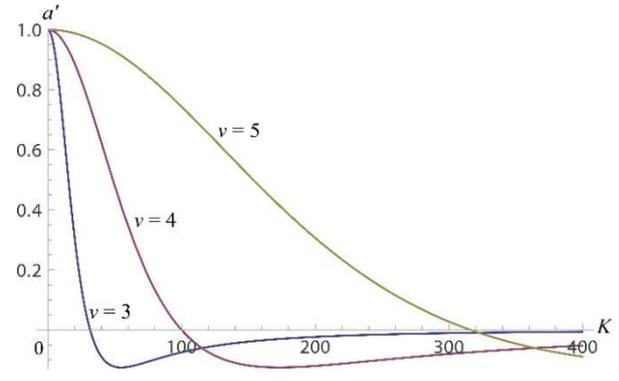

**Рис. 7**. Кривые ускорения для задачи (11) при различных $v$.

**Рис. 8**. Кривые производной ускорения для задачи (11) при различных $v$.

$$T_K = K(L+t_s) + t_w/K + K \cdot L + t_r + t_p, \quad (4)$$

что равносильно

$$T_K = 2L \cdot K + t_s \cdot K + t_r + t_p + t_w/K. \quad (5)$$

Приводя правую часть к общему знаменателю, получаем

$$T_K = \frac{K^2(2L+t_s) + K(t_r+t_p) + t_w}{K}. \quad (6)$$

Из формул (3) и (6) получаем следующую формулу для ускорения $a$:

$$a = \frac{T_1}{T_K} = \frac{K(2L+t_s+t_r+t_p+t_w)}{K^2(2L+t_s) + K(t_r+t_p) + t_w}. \quad (7)$$

Свяжем величины $t_s$ и $t_w$ через новую переменную $v$ следующим уравнением:

$$v = \lg(t_w/t_s). \quad (8)$$

Тогда

$$t_s = 10^{-v} t_w. \quad (9)$$

Подставляя значение $t_s$ из уравнения (9) в уравнение (7), получим

$$a = \frac{T_1}{T_K} = \frac{K(2L+10^{-v}t_w+t_r+t_p+t_w)}{K^2(2L+10^{-v}t_w) + K(t_r+t_p) + t_w}. \quad (10)$$

Исследуем, как для некоторой фиксированной задачи выглядят графики зависимости ускорения от числа рабочих. Пусть имеется некоторая задача с решением в пространстве $\mathbb{R}^n$. Предположим, что

$$n = 10^4, \ t_w = n^3 = 10^{12}, \ t_p = t_r = n = 10^4, \ L = 0.5. \quad (11)$$

Подставляя указанные значения в формулу (10), получаем

$$a(K) = \frac{K(1+10^{12-v}+2 \cdot 10^4 + 10^{12})}{K^2(1+10^{12-v}) + 2K \cdot 10^4 + 10^{12}} \approx \frac{K(10^{8-v}+2+10^8)}{10^{8-v}K^2 + 2K + 10^8}. \quad (12)$$

На рис. 7 приведены кривые ускорения $a$, вычисляемые по формуле (12) для различных значений параметра $v$.

Границами масштабируемости в каждом случае являются точки максимумов кривых ускорения, то есть – это точки, где производная ускорения равна нулю. Для определения таких точек вычислим производную по $K$ для ускорения, вычисляемого по формуле (7):

$$a'(K) = \frac{(2L + t_s + t_r + t_p + t_w)(t_w/K^2 - 2L - t_s)}{\left(K(2L + t_s) + t_r + t_p + t_w/K\right)^2}. \qquad (13)$$

Соответственно, для задачи (11), продифференцировав (12), получаем

$$a'(K) \approx \frac{(10^{8-\nu} + 2 + 10^8)\left(10^8 - 10^{8-\nu} K^2\right)}{(10^{8-\nu} K^2 + 2K + 10^8)^2}. \qquad (14)$$

Примеры графиков производных ускорения $a'$, вычисляемых по формуле (14), приведены на рис. 8.

Для того, чтобы вычислить нули производной (13), найдем корни уравнения

$$\frac{(2L + t_s + t_r + t_p + t_w)(t_w/K^2 - 2L - t_s)}{\left(K(2L + t_s) + t_r + t_p + t_w/K\right)^2} = 0. \qquad (15)$$

Поделив обе части уравнения (15) на $(2L + t_s + t_r + t_p + t_w)$, и умножив их на $\left(K(2L + t_s) + t_r + t_p + t_w/K\right)^2$, получим

$$t_w/K^2 - 2L - t_s = 0, \qquad (16)$$

что равносильно

$$\frac{K^2}{t_w} = \frac{1}{2L + t_s}, \qquad (17)$$

то есть

$$K = \sqrt{\frac{t_w}{2L + t_s}}. \qquad (18)$$

Таким образом, границы масштабируемости *BSF*-программы определяется следующим неравенством

$$K \leq \sqrt{\frac{t_w}{2L + t_s}}, \qquad (19)$$

где $K$ – количество узлов-рабочих; $t_w$ – время выполнения всех вычислений в рамках итерации бригадой из одного рабочего; $t_s$ – время, необходимое для посылки задания одному рабочему; $L$ – затраты на инициализацию операции передачи сообщения. Примечательно то, что границы масштабируемости *BSF*-программы не зависят от затрат на пересылку результатов от рабочих мастеру и от времени обработки этих результатов на узле-мастере. Однако, как будет показано ниже, эти параметры оказывают существенное влияние на эффективность распараллеливания.

Продемонстрируем, как полученная оценка может применяться на практике. Пусть имеется некоторая *BSF*-программа, для которой параметр $n$ (размерность задачи) характеризует объем исходных данных. Предположим, что затраты на посылку задания одному рабочему составляют $O(n)$, а суммарная временная стоимость вычислений, выполняемых рабочими, равна $O(n^3)$. Тогда по формуле (19) получаем $K \leq \sqrt{O(n^3)/O(n)}$, то есть $K \leq O(n)$. Это означает, что верхняя граница масштабируемости программы будет расти пропорционально росту размерности задачи, и, следовательно, мы можем характеризовать такую программу как *хорошо масштабируемую*.

Предположим теперь, что временная стоимость посылки задания одному рабочему по-прежнему составляет $O(n)$, а суммарная временная стоимость вычислений, выполняемых рабочими,

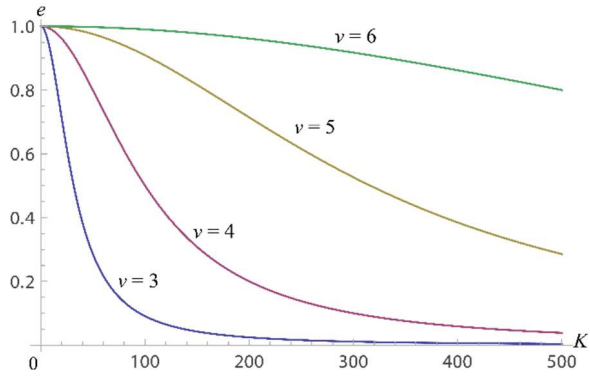
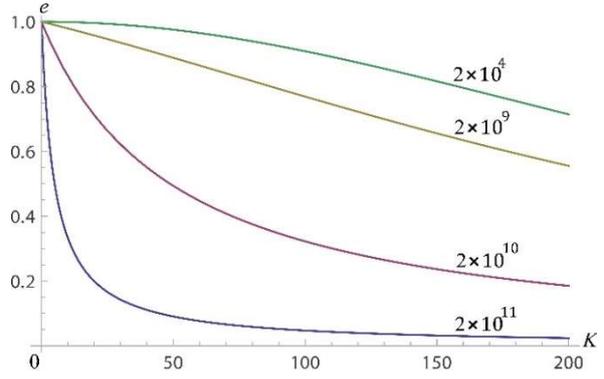

**Рис. 9**. Эффективность распараллеливания задачи (11) при различных $v$.

**Рис. 10**. Влияние $t_p + t_r$ на эффективность распараллеливания ($v = 5$).

равна $O(n^2)$. Тогда по формуле (19) получаем $K \leq \sqrt{O(n^2)/O(n)}$, то есть $K \leq \sqrt{O(n)}$. Это означает, что верхняя граница масштабируемости программы будет расти как корень квадратный от размерности задачи. Такую программу мы можем характеризовать как *ограниченно масштабируемую*.

В заключение рассмотрим случай, когда в рамках одной итерации временная стоимость посылки задания одному рабочему пропорциональна суммарной временной стоимости вычислений, выполняемых рабочими, и равна $O(n)$. В этом случае из формулы (19) получаем $K \leq \sqrt{O(n)/O(n)}$. Это означает, что верхняя граница масштабируемости программы ограничена некоторой константой, не зависящей от размерности задачи. Такую программу мы можем характеризовать как *плохо масштабируемую*.

Таким образом, мы можем сделать вывод, что *BSF*-программа будет обладать хорошей масштабируемостью, если временные затраты на посылку задания одному рабочему будут пропорциональны кубическому корню от суммарных временных затрат на решение задачи рабочими.

Оценим теперь эффективность $e$ распараллеливания *BSF*-программы. Используя (3) и (6), имеем

$$e = \frac{T_1}{K \cdot T_K} = \frac{2L + t_s + t_r + t_p + t_w}{K^2(2L + t_s) + K(t_r + t_p) + t_w} =$$

$$= \frac{2L + t_s}{K^2(2L + t_s) + K(t_r + t_p) + t_w} + \frac{t_r + t_p}{K^2(2L + t_s) + K(t_r + t_p) + t_w} + \frac{t_w}{K^2(2L + t_s) + K(t_r + t_p) + t_w}.$$

В предположении $K \gg 1$ отсюда следует, что

$$e \approx \frac{t_w}{K^2(2L + t_s) + K(t_r + t_p) + t_w}.$$

Поделив числитель и знаменатель на $t_w$, получаем итоговую формулу

$$e \approx \frac{1}{1 + \left(K^2(2L + t_s) + K(t_r + t_p)\right)/t_w}. \quad (20)$$

Указанная формула является асимптотически точной при $K \to \infty$. Подставляя значение $t_s$ из уравнения (9), можно получить следующий вариант формулы (20):

$$e \approx \frac{1}{1 + \left(K^2(2L + 10^{-v} t_w) + K(t_r + t_p)\right)/t_w}. \quad (21)$$

Подсчитаем по формуле (21) эффективность распараллеливания задачи (11):

$$e \approx \frac{1}{1+\left(K^2(1+10^{12-v})+2K\cdot 10^4\right)/10^{12}}. \quad (22)$$

На рис. 9 приведены графики эффективности распараллеливания задачи (11) для различных значений $v$, построенные с использованием формулы (22). Указанные графики показывают, что величина $v = \lg(t_w/t_s)$ оказывает существенное влияние на эффективность распараллеливания. Чем больше соотношение $t_w/t_s$, тем выше эффективность распараллеливания.

Сумма $t_r + t_p$ также оказывает существенное влияние на эффективность распараллеливания. Это можно увидеть на рис. 10, где приведены графики эффективности распараллеливания задачи (11) при различных значениях суммы $t_r + t_p$, указанных на кривых. Параметр $t_s$ в этом случае имеет фиксированное значение для всех графиков: $t_s = 10^{-v} t_w = 10^{-5} \cdot 10^{12} = 10^7$. Можно видеть, что при $t_r + t_p = 2\cdot 10^{11}$ эффективность распараллеливания на 20 процессорных узлах не превышает 20%, при этом, как показывает рис. 7, верхняя граница масштабируемости BSF-программы с такими параметрами лежит в районе 300 процессорных узлов.

## 6. Заключение

В работе описана новая модель параллельных вычислений *BSF* (Bulk Synchronous Farm) – блочно-синхронная ферма, ориентированная на вычислительные системы с массовым параллелизмом, включающие в себя сотни тысяч процессорных узлов и имеющие экзафлопный уровень производительности. *BSF*-компьютер представляет собой множество однородных процессорных узлов с приватной памятью, соединенных сетью, позволяющей передавать данные от одного процессорного узла другому. Среди процессорных узлов выделяется один, называемый мастером. Остальные $K$ узлов называются рабочими. *BSF*-компьютер работает по схеме *SPMD*. Описана структура *BSF*-программы. Построена стоимостная метрика для оценки времени выполнения *BSF*-программы. На основе предложенной стоимостной метрики получена оценка для верхней границы масштабируемости *BSF*-программ. Данная оценка позволяет сделать вывод, что *BSF*-программа будет обладать хорошей масштабируемостью, если временные затраты на посылку задания рабочему будут пропорциональны кубическому корню от суммарных временных затрат на решение задачи рабочими. Также получены формулы для оценки эффективности распараллеливания *BSF*-программ.

В рамках дальнейших исследований планируется решить следующие задачи:
1) разработать формализм для описания *BSF*-программ с использованием функций высшего порядка;
2) выполнить проектирование и реализацию каркаса для быстрой разработки *BSF*-программ на базе MPI (в виде библиотеки на языке C++);
3) провести вычислительные эксперименты на кластерной вычислительной системе с использованием искусственных и реальных задач для подтверждения адекватности модели *BSF*.

## Литература

# Analytical study of the "master-worker" framework scalability on multiprocessors with distributed memory

L.B. Sokolinsky

South Ural State University

The paper is devoted to an analytical study of the "master-worker" framework scalability on multiprocessors with distributed memory. A new model of parallel computations called *BSF* is proposed. The *BSF* model is based on *BSP* and *SPMD* models. The scope of *BSF* model is the compute-intensive applications. The architecture of *BSF*-computer is defined. The structure of *BSF*-program is described. A formal cost metric is provided. Using this metric, the upper scalability bounds of *BSF* programs on distributed memory multiprocessors are evaluated. The formulas for estimating the parallel efficiency of *BSF* programs also proposed.

*Keywords:* parallel programming, "master-worker" framework, *BSF* model, computational complexity, scalability, parallel efficiency, distributed memory multiprocessors.